\documentclass[preprint,11pt]{elsarticle}

\usepackage{lineno,hyperref}
\usepackage{comment}
\usepackage{graphicx}
\usepackage{amsmath}
\usepackage{multirow}
\usepackage{comment}
\usepackage{float}
\restylefloat{table}

\usepackage[table,xcdraw]{xcolor}

\usepackage{xcolor}

\modulolinenumbers[5]

\journal{Genomics}

\begin{document}

\begin{frontmatter}

\title{i6mA-CNN: a convolution based computational approach towards identification of DNA N6-methyladenine sites in rice genome}

\author{Ruhul Amin}
\author{Chowdhury Rafeed Rahman}
\cortext[mycorrespondingauthor]{Corresponding author}
\ead{rafeed@cse.uiu.ac.bd}

\author{Md. Sadrul Islam Toaha}
\author{Swakkhar Shatabda}
\cortext[mycorrespondingauthor]{Corresponding author}
\ead{swakkhar@cse.uiu.ac.bd}

\address{Department of Computer Science and Engineering, United International University, Dhaka, 1207, Bangladesh}

\begin{abstract}
DNA N6-methylation (6mA) in Adenine nucleotide is a post replication modification and is responsible for many biological functions. Experimental methods for genome wide 6mA site detection is an expensive and manual labour intensive process. Automated and accurate computational methods can help to identify 6mA sites in long genomes saving significant time and money. Our study develops a convolutional neural network based tool i6mA-CNN capable of identifying 6mA sites in the rice genome. Our model coordinates among multiple types of features such as PseAAC inspired customized feature vector, multiple one hot representations and dinucleotide physicochemical properties. It achieves area under the receiver operating characteristic curve of 0.98 with an overall accuracy of 0.94 using 5 fold cross validation on benchmark dataset. Finally, we evaluate our model on two other plant genome 6mA site identification datasets besides rice. Results suggest that our proposed tool is able to generalize its ability of 6mA site identification on plant genomes irrespective of plant species. Web tool for this research can be found at: \url{https://cutt.ly/Co6KuWG}. Supplementary data (benchmark dataset, independent test dataset, comparison purpose dataset, trained model, physicochemical property values, attention mechanism details for motif finding) are available at \textit{\url{https://cutt.ly/PpDdeDH}}.
\end{abstract}


\end{frontmatter}

%\linenumbers

\section{Introduction}
N4-methylcytosine, N6-methyladenine and 5-methylcytosine exist in DNA of different species \cite{def1}. DNA $N^{6}$-methylation is a part of epigenetics where methyl groups are transferred to DNA molecules \cite{def3}. It is one kind of post replication modification. Such modification has been identified in three kingdoms of life such as Bacteria, Archaea, and Eukaryotes \cite{def2}. $N^{6}$-methylation occurs when methyl group is transferred to the $6^{th}$ position of purine ring of Adenine nucleotide. Such modification is responsible for discrimination between original DNA strand and newly synthesized DNA strand, gene transcription regulation, replication and repair \cite{def4}. 

Laboratory based experimental methods such as sequencing (SMRT-seq, MeDIP-seq) \cite{motiv4}, methylated DNA Immunoprecipitation \cite{motiv3},
capillary electrophoresis and laser-induced fluorescence \cite{motiv2} have been proposed to identify $N^{6}$-methyladenine (6mA) sites in the genome. 6mA profile of rice has been obtained recently using mass spectrometry analysis and 6mA Immunoprecipitation followed by sequencing (IP-seq) \cite{motiv1}. Genome wide identification of 6mA sites is labor intensive and costly using such experimental methods. Computational experiment based pLogo plot (\cite{Res5}) on benchmark dataset regarding rice 6mA site identification shows nucleotide distribution to be different near 6mA and non-6mA sites \cite{Lit3}. So, accurate computational methods can be effective for automatic 6mA site identification in rice genome.

 iDNA6mA-PseKNC tool was established for $N^{6}$-methyladenine site identification on Mus musculus genome dataset containing 1934 samples in each class \cite{Lit1}. They used custom feature vector for each sample inspired by PseAAC \cite{motiv5}. They used LibSVM package 3.18 for their classification task. This same package was used in i6mA-Pred tool for the same task in rice genome dataset containing 880 samples in each class using similar feature vector \cite{Lit2}. They used Maximum Relevance Maximum Distance (MRMD) method (\cite{Lit21}) along with Incremental Feature Selection (IFS) for limiting feature space. This same dataset was used to train and build iDNA6mA tool which utilized one hot matrix as sample feature in \cite{Lit4}. They used sequential one dimensional convolutional neural network architecture for classification. This same approach was used in \cite{Lit7} to develop SNNRice6mA tool. The only difference was that they used group normalization layer in between convolution layers and an adaptive learning rate. The same benchmark rice genome dataset was used to build i6mA-DNCP tool \cite{Lit5}. They used di-nucleotide composition frequency and dinucleotide based DNA properties as sample features. TreeBagging algorithm was used as their classifier. 
They implemented heuristic based DNA property selection for selecting four dinucleotide based DNA properties for their classification. iDNA6mA-Rice tool was developed using a rice genome dataset containing 1,54,000 samples in each of the two classes  \cite{Lit3}. They used one hot vector based linear feature space with Random Forest algorithm. MM-6mAPred tool was developed in \cite{Lit6}, where neighbour dependency information was used to detect 6mA site in rice genome leveraging Markov Model. 6mA-Finder tool was developed in \cite{Lit8} using seven sequence oriented information and three physicochemical based features. They used Random Forest classifier on an optimal feature group using Recursive feature elimination (RFE) strategy.  

The main limitation of these methods is that they require single feature vector for training and validation purpose irrespective of the heterogeneity of the features they are using. Moreover, most of these proposed algorithms are unable to use two dimensional matrix oriented features which limits local and sequential pattern detection capability.  

We propose i6mA-CNN, a one dimensional CNN based branched classifier which can identify 6mA sites in plant genome. We work with four different types of feature matrices used for DNA sequence representation - PseAAC inspired customized feature matrix, monomer one hot matrix, dimer one hot matrix and dimer physicochemical properties. We use correlation based dimer physicochemical property selection technique in order to remove unnecessary and/ or redundant property features. Our proposed model is able to coordinate between these heterogeneous feature matrices and learn distinguishing features between 6mA and non-6mA sites successfully. Accurate experimental results on a large benchmark dataset and success on independent test datasets show the effectiveness of our proposed tool.     

\section{ Materials and Methods}

\subsection{Benchmark Dataset Construction}
Gene Expression Omnibus (GEO) database maintained by National Center for Biotechnology Information (NCBI) has a total of 2,65,290 6mA site containing sequences each of 41 base pair (bp) length with 6mA site at the center \cite{Data1}. CD-HIT program (\cite{Data2}) with 80\% (following similar research conducted in \cite{Lit3}) similarity threshold has been implemented on these sequences in order to avoid redundancy bias which resulted in 1,54,000 mA site containing sequences. This is our positive set of samples. Same number of negative samples (no 6mA site) have been obtained from NCBI to formulate a class balanced dataset following some specific criteria. Each 41 bp long negative sequence has Adenine at the center, where the center is not methylated (proven by experiment). 6mA sites have a tendency of occurring near GAGG rich sequences \cite{Data3}. In order to avoid such bias while classification, negative sequences rich with GAGG motifs were considered only to make negative data more objective.

\subsection{Sequence Representation}
A sample sequence of our dataset looks as follows:\\
$D = N_1, N_2, N_3, \cdots, N_L$\\
where, $L$ is the length of the DNA sequence, which is 41 in our case and $N_i \in \{A, T, C, G\}$.
With appropriate mathematical representation of these sequences, deep learning models can learn class distinguishing features from the sequence local and global patterns \cite{Seq1}. It is clear from Figure \ref{fig:model} that our model is capable of coordinating between multiple types of feature vectors and matrices using branch like structure which has been explained in detail in Subsection \ref{model}. 
We use the following sequence representation features in our model: \\

  \subsubsection{PseAAC Inspired Feature Representation}
  PseAAC (Pseudo Amino Acid Composition) has shown success in dealing with protein sequences \cite{Seq4, Seq5}. Inspired by such success, PseKNC (Pseudo K-tuple Nucleotide Composition) has been introduced for dealing with DNA/RNA sequences in computational biology \cite{Seq2, Seq3}. The general form of PseKNC has been designed in such a way that their values will depend on user specified feature extraction techniques from DNA samples. \\ \\
  The four types of nucleotides of DNA can be classified into three categories of attributes. These attributes along with representation code have been provided in Table \ref{table: encoding}. The categories are as follows:\\
  \textbf{Ring number:} A and G have two rings, whereas C and T have only one ring. \\
  \textbf{Chemical functionality:} A and C are from amino
group, whereas G and T are from keto group. \\
  \textbf{Hydrogen bonding:} C and G are bonded to each other with 3 hydrogen bonds (strong), whereas A and T with only two (weak). \\ \\
 The above mentioned properties have significant impact on biological functions as well \cite{Seq6}. The $i^{th}$ nucleotide $N_i$ of a sequence can be represented by $(x_i, y_i, z_i)$, where $x_i$, $y_i$ and $z_i$ represent ring structure, functional group and hydrogen bonding code (see Table \ref{table: encoding}) of nucleotide $N_i$,  respectively.
% \usepackage{multirow}
\begin{table}[b]
\begin{center}
\begin{tabular}{|c|c|c|c|}
\hline
\textbf{Category}                                                            & \textbf{Attribute}              & \textbf{Nucleotides} & \textbf{Code} \\ \hline
\multirow{2}{*}{\begin{tabular}[c]{@{}c@{}}Ring \\ Structure\end{tabular}}   & Purine                          & A, G                 & 1             \\ \cline{2-4} 
                                                                             & \multicolumn{1}{l|}{Pyrimidine} & C, T                 & 0             \\ \hline
\multirow{2}{*}{\begin{tabular}[c]{@{}c@{}}Functional \\ Group\end{tabular}} & Amino                           & A, C                 & 1             \\ \cline{2-4} 
                                                                             & Keto                            & G, T                 & 0             \\ \hline
\multirow{2}{*}{\begin{tabular}[c]{@{}c@{}}Hydrogen\\ Bonding\end{tabular}}  & Strong                          & C, G                 & 1             \\ \cline{2-4} 
                                                                             & Weak                            & A, T                 & 0             \\ \hline
\end{tabular}
\caption{Nucleotide categorical attribute-wise encoding}
\label{table: encoding}
\end{center}
\end{table}
 Besides these three properties, we consider lingering density of each nucleotide to preserve sequence oriented features.  
 Lingering density of nucleotide $N_{i}$ is defined below: \\ \\
$D_i = \frac{1}{L_i}\sum_{j=1}^{L_i} f(N_j)$ \\ \\
Here, $L_i$ is the length of the substring up to nucleotide $N_i$ and \\ \\
  $f(N_j) =
    \begin{cases}
      1, & \text{if $N_i = N_j$}\\
      0, & \text{otherwise}\\
    \end{cases} $ \\ \\
So, for each nucleotide, we have four properties to calculate. Since each sequence in our dataset is 41 bp long, we have a $41\times4$ dimensional feature matrix for each sequence. We illustrate this four property representation scheme on a toy example sequence "ACGGA" in Table \ref{table: PseAAC}. The resultant
matrix representation is:
$[[1,1,0,1],[0,1,1,0.5],[1,0,1,0.33],[1,0,1,0.5],[1,1,0,0.4]]$
\begin{table}[b]
\begin{center}
\begin{tabular}{|c|c|c|c|c|c|}
\hline
\textbf{Pos} & \textbf{Nuc} & \textbf{X} & \textbf{Y} & \textbf{Z} & \textbf{D} \\ \hline
1            & A            & 1          & 1          & 0          & 1/1 = 1    \\ \hline
2            & C            & 0          & 1          & 1          & 1/2 = 0.5  \\ \hline
3            & G            & 1          & 0          & 1          & 1/3 = 0.33 \\ \hline
4            & G            & 1          & 0          & 1          & 2/4 = 0.5  \\ \hline
5            & A            & 1          & 1          & 0          & 2/5 = 0.4  \\ \hline
\end{tabular}
\caption{Four property representation scheme detail of "ACGGA" sequence}
\label{table: PseAAC}
\end{center}
\end{table}
   
 \subsubsection{Monomer One Hot Representation}
    Tool iDNA6mA was developed based on monomer one hot encoding \cite{Lit4}. Their one dimensional convolutional neural network (1D CNN) was quite successful in class discrimination based on one hot encoding based mathematical representation. Each branch of our designed model is also 1D CNN oriented. So, we have used this particular feature. 
   There are four types of nucleotides in DNA sequence and each of them is represented by a four size one hot vector. A, T, C and G are represented by (1,0,0,0), (0,1,0,0), (0,0,1,0) and (0,0,0,1),  respectively. So, a 41 bp long sequence is represented by a $41\times4$ dimensional matrix.
   
 \subsubsection{Dimer one hot representation}
    There is statistically significant difference (considering p-value of 0.05) between 6mA and non-6mA site containing samples in terms of dinucleotide composition frequencies \cite{Lit5}. There are sixteen possible unique dimers such as AA, AT, AC, \ldots in DNA sequences. We represent each dimer by a 16 size one hot vector. 1D CNN based feature extraction technique is able to learn from class distinguishing dimer patterns which include difference in frequency. In a 41 length sequence, there are 40 overlapping dimers such as $N_1N_2$, $N_2N_3$, $N_3N_4$, \ldots. So, each sample of our dataset is represented by a $40\times16$ dimensional matrix.
    
 \subsubsection{Dimer Physicochemical Property Based Representation}
     According to \cite{Seq7}, Adenine methylation may have impact on DNA structure and its stability. We have used the 90 dinucleotide physicochemical properties used by PseKNC tool \cite{Seq2}. So, each dimer of our sample sequence is represented by a 90 size Z-score normalized vector. As a result, each 41 bp long sample has been represented by a $40\times90$ dimensional matrix.  \\ \\
     This is certainly a large matrix. If the dinucleotide physicochemical property set contains features which include redundant or less important information in terms of our classification task, it will lead to overfitting problem \cite{Seq8}. The steps of dinucleotide physicochemical property (DPP) selection are as follows:
     \begin{itemize}
         \item We use Logistic regression based model for 6mA site identification task using one DPP at a time.
         \item We note the 5 fold cross validation Mathew's correlation coefficient (MCC) score for each DPP and select the the top 20 DPPs according to obtained MCC score. The higher the MCC, the better is the DPP for our identification task. It is to note that MCC score is better than F1 score and accuracy for binary classification evaluation \cite{Seq9}.
         \item We select the topmost DPP. Now we want to remove redundant information. Each DPP is represented by a 16 size vector as there are 16 possible dimers. DPPs which have a high positive or negative correlation coefficient represent the same type of information. We calculate Pearson's correlation coefficient for each of the 19 other selected DPPs with the topmost DPP. 
         \item We leave out those DPPs which have correlation score greater than 0.9 (positive correlation) or less than -0.9 (negative correlation) with the topmost DPP. 
     \end{itemize}
     Thus we select nine DPPs out of 90 which include \textit{Roll rise, major Groove distance, twist stiffness, protein induced deformability, mobility to bend towards major groove, melting temperature, Hartman trans free energy, protein DNA twist and tip.}

\begin{figure}[!htb]
\begin{center}
\includegraphics[width=0.9\columnwidth]{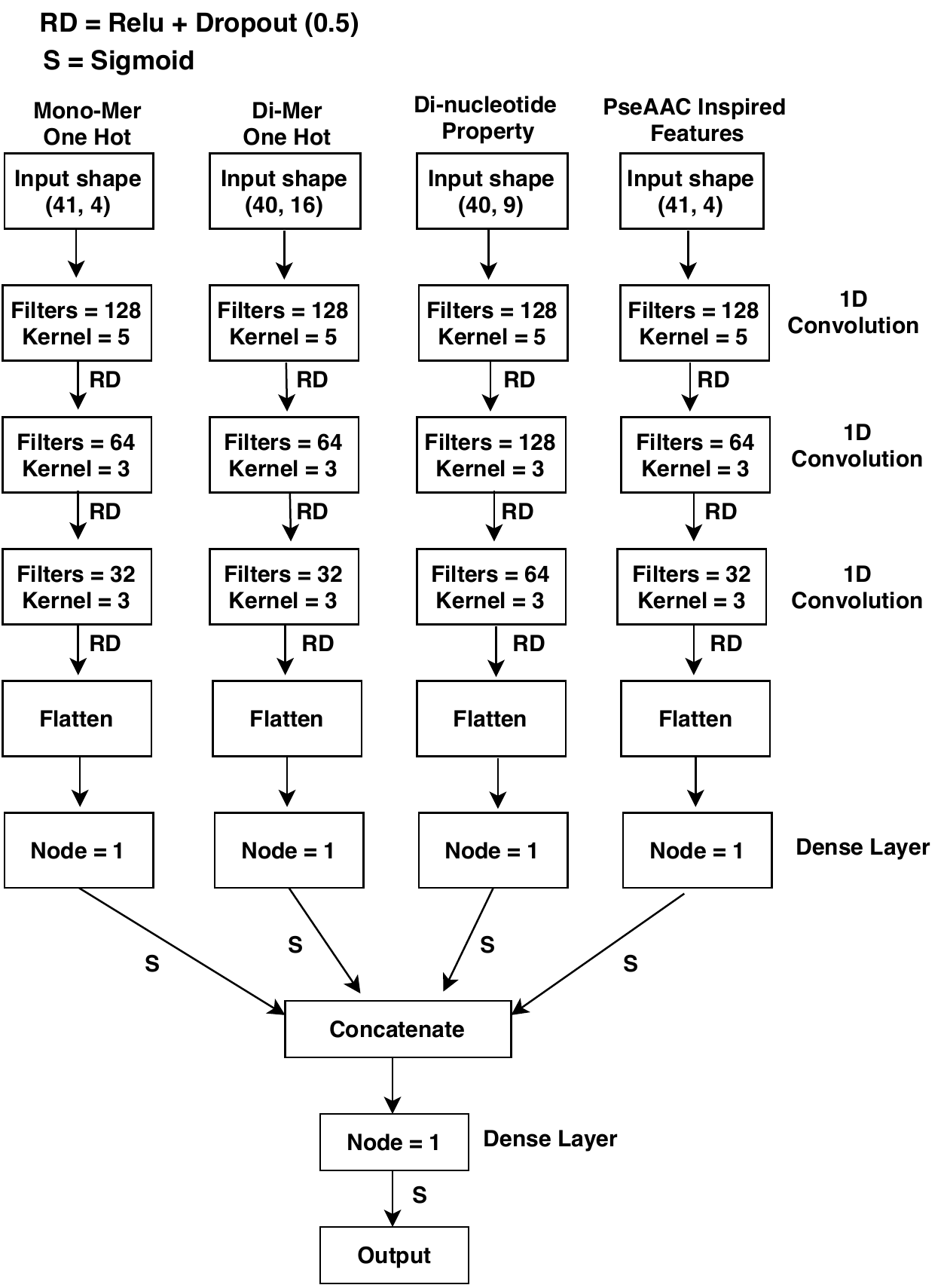}
\end{center}
\caption{Proposed Model Architecture}
\label{fig:model}
\end{figure}

\subsection{Model Architecture} \label{model}
Our model architecture has been shown in Figure \ref{fig:model}. There are four branches for four different feature matrices (described in detail in Subsection \textbf{Sequence Representation}). Each branch has three 1D convolution layers put sequentially. The kernel size and filter number of each convolution layer have been shown in the figure. 1D CNN plays important role in position independent local feature extraction \cite{Mod1, Mod2}. Each branch of the architecture learns class distinguishing features regarding its corresponding input feature matrix independent of the other three branches. The monomer one hot representation of the leftmost branch represents raw representation of sample sequence. Model parameters of this branch learn from this raw representation free from any feature specific knowledge bias. This branch has the potential of learning those class specific hidden features which may not be apparent to humans. The second branch which has dimer one hot representation as input helps the model learn statistically significant differences between 6mA and non-6mA site containing samples in terms of dinucleotide pattern and frequency of occurrence. Such difference has been pointed out in \cite{Lit5}. The third branch uses dimer physicochemical property based representation. These properties played an important role in DNA specific classification and identification tasks such as promoter classification \cite{Mod2}, 6mA site identification \cite{Lit5}, recombination spot identification \cite{Mod3} and DNase I hypersensitive sites identification \cite{Mod4}. This third branch facilitates learning the interaction between these property values in a sequence order basis. The last branch is based on learning from ring structure, functional group, hydrogen bonding and lingering density based nucleotide features. These features have impact on DNA biological functions \cite{Seq6, Mod5}. This last branch parameters learn to identify patterns from these sequence coupled features.   

We need to have a way to combine these branches for a mutual classification decision depending on the four heterogeneous feature matrices. The convolutional part of each branch i returns a matrix of dimension $m_i \times n_i$. We flatten and convert this matrix into a  linear vector. We pass this vector through a dense node having Sigmoid activation function, which outputs a value in the range of 0 to 1. This single value obtained from the $i^{th}$ branch is its representative for 6mA identification task. We have total four values in a range of 0 to 1 from the four branches, which we concatenate and turn into a four size vector. This four size vector is the representative vector of all four branches. The final dense layer with Sigmoid activation function provides the output value depending on the four values of the concatenated vector. If the output value is greater than 0.5, then we consider the sample sequence to have 6mA site. Else, we decide that there is no 6 mA site in the input sequence.

\subsection{Performance Evaluation and Model Selection}
Accuracy (Acc), sensitivity (Sn), specificity (Sp) and Mathew's correlation coefficient (MCC) have been used as evaluation metrics which are described as follows: \\ \\
$Acc = \dfrac{TP + TN}{TP + TN + FP + FN}$\\ \\
$Sn = \dfrac{TP}{TP + FN}$ \\ \\
$Sp = \dfrac{TN}{TN + FP}$ \\ \\
$MCC = \dfrac{TP \times TN - FP \times FN}{\sqrt{(TP + FP)(TP + FN)(TN + FP)(TN + FN)}}$ \\ \\
Here, TP, FP, TN and FN represent true positive, false positive, true negative and false negative, respectively. We consider the 6mA site containing samples as positive class samples.
Apart from the above four metrics, we have also used area under receiver operating characteristic (auROC) curve for performance evaluation. This metric provides classifier performance irrespective of provided threshold for class discrimination to the classifier. This score has an impact on classification confidence value for each sample as well. Acc, Sn, Sp and auROC lie in the range [0, 1] while for MCC score, the range is from -1 to 1. Higher value indicates better classification ability.

We have tuned and selected our model hyperparameters using 5 fold cross validation on our benchmark dataset. We took a local search oriented approach where we went in uphill direction in terms of obtained MCC value. The tuned model hyperparameters are as follows:
\begin{itemize}
  \item \textbf{Convolution layer:} Change in number of convolution layers in each branch, convolution filter number and kernel size in each convolution layer
  \item \textbf{Learning rate:} Experimentation with different learning rates such as 0.01, 0.001, 0.0001, 0.00001 and adaptive learning rate scheme
  \item \textbf{Dropout rate:} Testing out dropout rate of 0.1, 0.3 and 0.5.
  \item \textbf{Optimizer:} Model training with adaptive moment estimation (ADAM), stochastic gradient descent (SGD) and Nesterov accelerated gradient (NAG) optimizer. 
  \item \textbf{Activation function:} Experimentation with Relu, Tanh and LeakyRelu activation function in initial convolution layers
\end{itemize}

As shown in Figure \ref{fig:model}, Each branch has three convolution layers. These three consecutive layers contain kernel size of 5, 3 and 3, whereas the number of filters are 128, 64 and 32, respectively. Learning rate of 0.0001 has been used in order to avoid divergence while training.  Dropout of 0.5 has been used for overfitting avoidance. Addition of dropout changes network connectivity on each training epoch and thus helps model to understand the training sample features instead of memorizing them \cite{Mod7}. ADAM has been chosen as the optimizer for model weight update during training. Each intermediate convolution layer output goes through Relu activation function which is very popular for being effective and simple \cite{Mod6}.

\section{Results and Discussion}
i6mA-Pred \cite{Lit2}, iDNA6mA-Rice \cite{Lit3}, iDNA6mA \cite{Lit4}, i6mA-DNCP \cite{Lit5}, MM-6mAPred \cite{Lit6}, SNNRice6mA \cite{Lit7} and 6mA-Finder \cite{Lit8} are some of  the state-of-art tools which can identify 6mA sites in rice genome. For comparison purpose of our proposed tool with these state-of-the-art tools, consistency in dataset usage and evaluation method are necessary.

\begin{figure}[!htb]
\begin{center}
\includegraphics[width=1\columnwidth]{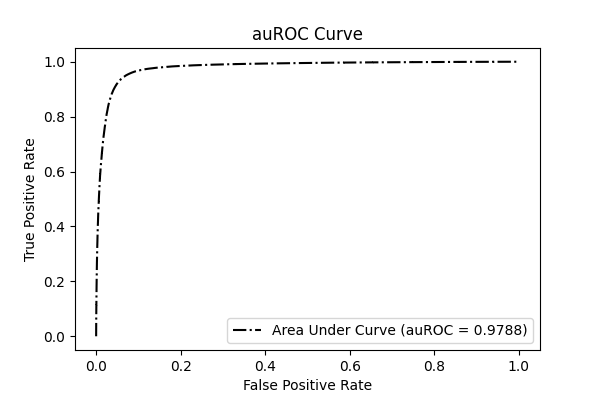}
\end{center}
\caption{ROC curve of i6mA-CNN on benchmark dataset}
\label{fig: ROC}
\end{figure}

The benchmark dataset that we have used for this research has also been used for constructing the latest rice genome i6mA site identification tool \textbf{iDNA6mA-Rice} and SNNRice6mA \cite{Lit3, Lit7}. Tools for the same task have been developed in \cite{Lit2, Lit4, Lit5, Lit6, Lit8}, though they used a different dataset for benchmarking which contains 880 samples in each of the two classes. They used either Jackknife testing or 10 fold cross validation for algorithm and model selection. For comparison purpose, we also test our custom model on the same 880 sample per class dataset using similar validation methods. This comparison purpose dataset has been obtained from NCBI GEO under accession number GSE103145. All sequences with 6mA site in the center and with modification score greater than 30 (according to Methylome Analysis Technical Note, 30 is the minimum score to call a nucleotide as modified) have been considered as positive samples. Each sample sequence is 41 bp long. CD-HIT (\cite{Data2}) with 60\% similarity threshold has been applied on the chosen positive sequences in order to remove redundancy. Thus the 880 positive samples were obtained. The 880 negative samples of this dataset have been obtained in a manner similar to our negative benchmark dataset samples.
Detailed performance comparison between the tools used for 6mA site identification has been provided in Table \ref{table: comparison}. All the tool results except for our proposed tool i6mA-CNN have been obtained from corresponding research articles (\cite{Lit2, Lit3, Lit4, Lit5, Lit6, Lit7, Lit8}). Our proposed tool clearly outperforms all other tools for similar task by all five performance metrics. Tool i6mA-CNN achieves a high 5 fold cross validation auROC score of 0.98 on our benchmark dataset. ROC curve has been shown in Figure \ref{fig: ROC}.

% \usepackage{multirow}
\begin{table}[]
\begin{tabular}{|c|c|c|c|c|c|c|c|}
\hline
\textbf{Dataset}            & \textbf{\begin{tabular}[c]{@{}c@{}}Evaluation \\ Method\end{tabular}}                    & \textbf{Tool}                                          & \multicolumn{1}{l|}{\textbf{Sn}} & \multicolumn{1}{l|}{\textbf{Sp}}                    & \multicolumn{1}{l|}{\textbf{Acc}} & \multicolumn{1}{l|}{\textbf{MCC}} & \textbf{auROC} \\ \hline
                            &                                                                                          & i6mA-Pred                                              & 83.41\%                          & 83.64\%                                             & 83.52\%                           & 0.67                              & 0.91           \\ \cline{3-8} 
                            & \multirow{-2}{*}{\begin{tabular}[c]{@{}c@{}}Jackknife\\  Testing\end{tabular}}           & \textbf{i6mA-CNN}                                      & \textbf{90.61\%}                 & \textbf{94.12\%}                                    & \textbf{93.15\%}                  & \textbf{0.85}                     & \textbf{0.98}  \\ \cline{2-8} 
                            &                                                                                          & iDNA6mA                                                & 86.7\%                           & 86.59\%                                             & 86.64\%                           & 0.73                              & 0.93           \\ \cline{3-8} 
                            &                                                                                          & i6mA-DNCP                                              & 84.09\%                          & 88.07\%                                             & 86.08\%                           & 0.72                              & 0.93           \\ \cline{3-8} 
                            &                                                                                          & \multicolumn{1}{l|}{{\color[HTML]{242021} MM-6mAPred}} & 89.32\%                          & \multicolumn{1}{l|}{{\color[HTML]{242021} 90.11\%}} & 89.72\%                           & 0.79                              & \_             \\ \cline{3-8} 
                            &                                                                                          & 6mA-Finder                                             & \_                               & \_                                                  & \_                                & \_                                & 0.94           \\ \cline{3-8} 
\multirow{-7}{*}{Dataset 1} & \multirow{-5}{*}{\begin{tabular}[c]{@{}c@{}}10 Fold \\ Cross \\ Validation\end{tabular}} & \textbf{i6mA-CNN}                                      & \textbf{90.35\%}                 & \textbf{94.62\%}                                    & \textbf{92.48\%}                  & \textbf{0.85}                     & \textbf{0.98}  \\ \hline
                            &                                                                                          & iDNA6mA-Rice                                           & 93\%                             & 90.5\%                                              & 91.7\%                            & 0.83                              & 0.96           \\ \cline{3-8} 
                            &                                                                                          & \multicolumn{1}{l|}{{\color[HTML]{242021} SNNRice6mA}} & 94.33\%                          & 89.75\%                                             & 92.04\%                           & 0.84                              & 0.97           \\ \cline{3-8} 
\multirow{-3}{*}{Dataset 2} & \multirow{-3}{*}{\begin{tabular}[c]{@{}c@{}}5 Fold \\ Cross\\  Validation\end{tabular}}  & \textbf{i6mA-CNN}                                      & \textbf{95.13\%}                 & \textbf{92.81\%}                                    & \textbf{93.97\%}                  & \textbf{0.88}                     & \textbf{0.98}  \\ \hline
\end{tabular}
\caption{i6mA-CNN performance comparison with state-of-the-art tools. Best performance results have been marked in bold. \textbf{Dataset 1} is the 880 sample per class dataset used for comparison purpose, while \textbf{Dataset 2} is our benchmark dataset with 1,54,000 samples in each class}
\label{table: comparison}
\end{table}

Independent test set performance of a tool can help in proving the generalization ability of that tool. Following procedures similar to our benchmark dataset construction, we have constructed two independent test datasets in order to evaluate the robustness of our proposed tool in terms of $N^{6}$-Methyladenine site identification in plant genome. Fragaria Vesca (Wild strawberry) and Arabidopsis thaliana (cress weed plant) 6mA site containing sequences have been obtained from the MDR database \cite{Res1} and from NCBI GEO with accession number GSE81597 \cite{Res2}, respectively. Performance of i6mA-CNN on these datasets have been shown in detail in Table \ref{table: independent}. Results show the robustness of the proposed tool for 6mA site identification in plant genome.  

\begin{table*}[bth]
\begin{center}
\begin{tabular}{|c|c|c|c|c|}
\hline
\textbf{\begin{tabular}[c]{@{}c@{}}Independent Test\\ Dataset\end{tabular}} & \textbf{\begin{tabular}[c]{@{}c@{}}Total \\ Sample No.\end{tabular}} & \textbf{\begin{tabular}[c]{@{}c@{}}TP\\ Sample No.\end{tabular}} & \textbf{\begin{tabular}[c]{@{}c@{}}FN\\ Sample No.\end{tabular}} & \textbf{\begin{tabular}[c]{@{}c@{}}Success\\ Rate\end{tabular}} \\ \hline
Fragaria Vesca                                                              & 8979                                                                 & 8642                                                             & 337                                                              & 96.25\%                                                         \\ \hline
Arabidopsis Thaliana                                                        & 67650                                                                & 63252                                                            & 4398                                                             & 93.5\%                                                          \\ \hline
\end{tabular}
\caption{i6mA-CNN performance on independent test datasets}
\label{table: independent}
\end{center}
\end{table*}

\begin{comment}
Our tool has also been tested on Mus musculus (house mouse) genome dataset. It was used for constructing iDNA6mA-PseKNC tool \cite{Lit1}. The 6mA site containing positive samples were extracted from MethSMRT database (\cite{Res3}). Our tool achieved less than 50\% test accuracy on this particular dataset. This proves that 6mA site containing region pattern of plants are different from that of animals.
\end{comment}

Our model is a CNN based branched model with no time stamp oriented sequential information processing capability.
There are reasons why we did not use recurrent layers which utilize time stamp information while working with sequence type data. First, it is not possible to provide multiple feature representations simultaneously in each time stamp input in recurrent neural networks. Second, each time stamp of such networks only accept linear vector representation which would limit our current ability of using matrices. Third, recurrent neural networks such as RNN, GRU or LSTM give importance to positional information \cite{Res4}. But 
classification of genome related sample sequence often depends on the presence of potential motif type patterns irrespective of their position in the sequence \cite{Data3}. Finally, training and validation using recurrent layer based networks are costly because of sequential type of computation. 1D CNN based model having multiple convolution layers have the capability of encoding and learning both short and long range patterns. Parallel processing is possible in such networks which allow fast computation. Our customized model architecture is able to work with multiple heterogeneous features simultaneously and is able to utilize the benefits of feature matrix representation. Because of such inherent characteristics, our proposed tool is fast, accurate and shows performance superior to the remaining state-of-the-art tools for 6mA site identification in rice genome and also generalizes well to other plant genomes for the same task.     

\begin{table}[htb]
\begin{center}
\begin{tabular}{|c|c|}
\hline
\multicolumn{1}{|l|}{\textbf{Motif Sequence}} & \multicolumn{1}{l|}{\textbf{Active Occurrence \%}} \\ \hline
ATAT                                          & 5.2                                                \\ \hline
AGGC                                          & 4.1                                                \\ \hline
AAAA                                          & 4.0                                                \\ \hline
AAAT                                          & 3.9                                                \\ \hline
GAGG                                          & 3.5                                                \\ \hline
\end{tabular}
\caption{Potential motifs for $N^{6}$-methyladenine site identification}
\label{table: motif}
\end{center}
\end{table}

 We have found potential motifs which our deep learning based model actively searches for in order to perform 6mA site identification in rice genome. Motifs are class specific substrings found frequently in the sample sequences of that class. Following the procedure shown in \cite{Mod2}, we have customized an attention model (details of this model have been provided as supplementary information) based on our proposed CNN model provided in Figure \ref{fig:model} with a view to identifying the potential motifs. A list of potential motifs found for 6mA site identification on our benchmark dataset have been provided in Table~\ref{table: motif}. The top motif that we have found is \textbf{ATAT}. It was found to be active 5.2\% of the time during 6mA site identification. The pLogo plot constructed in \cite{Lit3} showed the significant difference of nucleotide distribution in postive and negative samples. This analysis has not been repeated in our current work. 

\section{Conclusion}
Tool i6mA-CNN is the outcome of the current research aimed at 6mA site identification in rice genome. The combination of four different feature matrices and redundant feature reduction have facilitated our customized CNN based architecture to be robust and accurate. Experimental results on two rice genome datasets and two other independent plant genome datasets related to 6mA site identification show the effectiveness of our tool. Proposed i6mA-CNN is expected to be an effective tool for automation in epigenetics related research. Future research may aim at constructing a tool that can perform 6mA site identification in both plant and animal genome effectively. 

\bibliography{main}

\begin{thebibliography}{10}
\expandafter\ifx\csname url\endcsname\relax
  \def\url#1{\texttt{#1}}\fi
\expandafter\ifx\csname urlprefix\endcsname\relax\def\urlprefix{URL }\fi
\expandafter\ifx\csname href\endcsname\relax
  \def\href#1#2{#2} \def\path#1{#1}\fi

\bibitem{def1}
P.~Feng, H.~Yang, H.~Ding, H.~Lin, W.~Chen, K.-C. Chou, idna6ma-pseknc:
  Identifying dna n6-methyladenosine sites by incorporating nucleotide
  physicochemical properties into pseknc, Genomics 111~(1) (2019) 96--102.

\bibitem{def3}
F.~von Meyenn, M.~Iurlaro, E.~Habibi, N.~Q. Liu, A.~Salehzadeh-Yazdi,
  F.~Santos, E.~Petrini, I.~Milagre, M.~Yu, Z.~Xie, et~al., Impairment of dna
  methylation maintenance is the main cause of global demethylation in naive
  embryonic stem cells, Molecular cell 62~(6) (2016) 848--861.

\bibitem{def2}
Z.~K. O’Brown, E.~L. Greer, N6-methyladenine: a conserved and dynamic dna
  mark, in: DNA Methyltransferases-Role and Function, Springer, 2016, pp.
  213--246.

\bibitem{def4}
D.~Wion, J.~Casades{\'u}s, N 6-methyl-adenine: an epigenetic signal for
  dna--protein interactions, Nature Reviews Microbiology 4~(3) (2006) 183--192.

\bibitem{motiv4}
B.~A. Flusberg, D.~R. Webster, J.~H. Lee, K.~J. Travers, E.~C. Olivares, T.~A.
  Clark, J.~Korlach, S.~W. Turner, Direct detection of dna methylation during
  single-molecule, real-time sequencing, Nature methods 7~(6) (2010) 461.

\bibitem{motiv3}
K.~R. Pomraning, K.~M. Smith, M.~Freitag, Genome-wide high throughput analysis
  of dna methylation in eukaryotes, Methods 47~(3) (2009) 142--150.

\bibitem{motiv2}
A.~M. Krais, M.~G. Cornelius, H.~H. Schmeiser, Genomic n6-methyladenine
  determination by mekc with lif, Electrophoresis 31~(21) (2010) 3548--3551.

\bibitem{motiv1}
C.~Zhou, C.~Wang, H.~Liu, Q.~Zhou, Q.~Liu, Y.~Guo, T.~Peng, J.~Song, J.~Zhang,
  L.~Chen, et~al., Identification and analysis of adenine n 6-methylation sites
  in the rice genome, Nature plants 4~(8) (2018) 554--563.

\bibitem{Res5}
J.~P. O'shea, M.~F. Chou, S.~A. Quader, J.~K. Ryan, G.~M. Church, D.~Schwartz,
  plogo: a probabilistic approach to visualizing sequence motifs, Nature
  methods 10~(12) (2013) 1211--1212.

\bibitem{Lit3}
L.~Hao, F.-Y. Dao, Z.-X. Guan, D.~Zhang, J.-X. Tan, Y.~Zhang, W.~Chen, H.~Lin,
  idna6ma-rice: a computational tool for detecting n6-methyladenine sites in
  rice, Frontiers in genetics 10 (2019) 793.

\bibitem{Lit1}
P.~Feng, H.~Yang, H.~Ding, H.~Lin, W.~Chen, K.-C. Chou, idna6ma-pseknc:
  Identifying dna n6-methyladenosine sites by incorporating nucleotide
  physicochemical properties into pseknc, Genomics 111~(1) (2019) 96--102.

\bibitem{motiv5}
H.-B. Shen, K.-C. Chou, Pseaac: a flexible web server for generating various
  kinds of protein pseudo amino acid composition, Analytical biochemistry
  373~(2) (2008) 386--388.

\bibitem{Lit2}
W.~Chen, H.~Lv, F.~Nie, H.~Lin, i6ma-pred: Identifying dna n6-methyladenine
  sites in the rice genome, Bioinformatics 35~(16) (2019) 2796--2800.

\bibitem{Lit21}
W.~Chen, P.~Feng, H.~Ding, H.~Lin, Classifying included and excluded exons in
  exon skipping event using histone modifications, Frontiers in genetics 9
  (2018) 433.

\bibitem{Lit4}
M.~Tahir, H.~Tayara, K.~T. Chong, idna6ma (5-step rule): Identification of dna
  n6-methyladenine sites in the rice genome by intelligent computational model
  via chou's 5-step rule, Chemometrics and Intelligent Laboratory Systems 189
  (2019) 96--101.

\bibitem{Lit7}
H.~Yu, Z.~Dai, Snnrice6ma: a deep learning method for predicting dna
  n6-methyladenine sites in rice genome, Frontiers in genetics 10 (2019) 1071.

\bibitem{Lit5}
L.~Kong, L.~Zhang, i6ma-dncp: computational identification of dna
  n6-methyladenine sites in the rice genome using optimized dinucleotide-based
  features, Genes 10~(10) (2019) 828.

\bibitem{Lit6}
C.~Pian, G.~Zhang, F.~Li, X.~Fan, Mm-6mapred: identifying dna n6-methyladenine
  sites based on markov model, Bioinformatics 36~(2) (2020) 388--392.

\bibitem{Lit8}
H.~Xu, R.~Hu, P.~Jia, Z.~Zhao, 6ma-finder: a novel online tool for predicting
  dna n6-methyladenine sites in genomes, Bioinformatics 36~(10) (2020)
  3257--3259.

\bibitem{Data1}
C.~Long, W.~Li, P.~Liang, S.~Liu, Y.~Zuo, Transcriptome comparisons of
  multi-species identify differential genome activation of mammals
  embryogenesis, Ieee Access 7 (2018) 7794--7802.

\bibitem{Data2}
W.~Li, A.~Godzik, Cd-hit: a fast program for clustering and comparing large
  sets of protein or nucleotide sequences, Bioinformatics 22~(13) (2006)
  1658--1659.

\bibitem{Data3}
C.~Zhou, C.~Wang, H.~Liu, Q.~Zhou, Q.~Liu, Y.~Guo, T.~Peng, J.~Song, J.~Zhang,
  L.~Chen, et~al., Identification and analysis of adenine n 6-methylation sites
  in the rice genome, Nature plants 4~(8) (2018) 554--563.

\bibitem{Seq1}
R.~K. Umarov, V.~V. Solovyev, Recognition of prokaryotic and eukaryotic
  promoters using convolutional deep learning neural networks, PloS one 12~(2)
  (2017) e0171410.

\bibitem{Seq4}
W.-Z. Zhong, S.-F. Zhou, Molecular science for drug development and biomedicine
  (2014).

\bibitem{Seq5}
G.-P. Zhou, W.-Z. Zhong, Perspectives in medicinal chemistry., Current topics
  in medicinal chemistry 16~(4) (2016) 381.

\bibitem{Seq2}
W.~Chen, T.-Y. Lei, D.-C. Jin, H.~Lin, K.-C. Chou, Pseknc: a flexible web
  server for generating pseudo k-tuple nucleotide composition, Analytical
  biochemistry 456 (2014) 53--60.

\bibitem{Seq3}
W.~Chen, H.~Lin, K.-C. Chou, Pseudo nucleotide composition or pseknc: an
  effective formulation for analyzing genomic sequences, Molecular BioSystems
  11~(10) (2015) 2620--2634.

\bibitem{Seq6}
K.-C. Chou, Low-frequency collective motion in biomacromolecules and its
  biological functions, Biophysical chemistry 30~(1) (1988) 3--48.

\bibitem{Seq7}
S.~Cheng, G.~Herman, P.~Modrich, Extent of equilibrium perturbation of the dna
  helix upon enzymatic methylation of adenine residues., Journal of Biological
  Chemistry 260~(1) (1985) 191--194.

\bibitem{Seq8}
D.~M. Hawkins, The problem of overfitting, Journal of chemical information and
  computer sciences 44~(1) (2004) 1--12.

\bibitem{Seq9}
D.~Chicco, G.~Jurman, The advantages of the matthews correlation coefficient
  (mcc) over f1 score and accuracy in binary classification evaluation, BMC
  genomics 21~(1) (2020) 6.

\bibitem{Mod1}
T.~Chen, R.~Xu, Y.~He, X.~Wang, Improving sentiment analysis via sentence type
  classification using bilstm-crf and cnn, Expert Systems with Applications 72
  (2017) 221--230.

\bibitem{Mod2}
R.~Amin, C.~R. Rahman, S.~Ahmed, M.~Sifat, H.~Rahman, M.~N.~K. Liton,
  M.~Rahman, M.~Khan, Z.~Hossain, S.~Shatabda, et~al., ipromoter-bncnn: a novel
  branched cnn based predictor for identifying and classifying sigma promoters,
  Bioinformatics (2019).

\bibitem{Mod3}
L.~Zhang, L.~Kong, irspot-pdi: Identification of recombination spots by
  incorporating dinucleotide property diversity information into chou's pseudo
  components, Genomics 111~(3) (2019) 457--464.

\bibitem{Mod4}
S.~Zhang, J.~Lin, L.~Su, Z.~Zhou, pdhs-dset: Prediction of dnase i
  hypersensitive sites in plant genome using ds evidence theory, Analytical
  biochemistry 564 (2019) 54--63.

\bibitem{Mod5}
K.-C. Chou, B.~Mao, Collective motion in dna and its role in drug
  intercalation, Biopolymers: Original Research on Biomolecules 27~(11) (1988)
  1795--1815.

\bibitem{Mod7}
N.~Srivastava, G.~Hinton, A.~Krizhevsky, I.~Sutskever, R.~Salakhutdinov,
  Dropout: a simple way to prevent neural networks from overfitting, The
  journal of machine learning research 15~(1) (2014) 1929--1958.

\bibitem{Mod6}
A.~F. Agarap, Deep learning using rectified linear units (relu), arXiv preprint
  arXiv:1803.08375 (2018).

\bibitem{Res1}
Z.-Y. Liu, J.-F. Xing, W.~Chen, M.-W. Luan, R.~Xie, J.~Huang, S.-Q. Xie, C.-L.
  Xiao, Mdr: an integrative dna n6-methyladenine and n4-methylcytosine
  modification database for rosaceae, Horticulture research 6~(1) (2019) 1--7.

\bibitem{Res2}
Z.~Liang, L.~Shen, X.~Cui, S.~Bao, Y.~Geng, G.~Yu, F.~Liang, S.~Xie, T.~Lu,
  X.~Gu, et~al., Dna n6-adenine methylation in arabidopsis thaliana,
  Developmental cell 45~(3) (2018) 406--416.

\bibitem{Res4}
W.~Yin, K.~Kann, M.~Yu, H.~Sch{\"u}tze, Comparative study of cnn and rnn for
  natural language processing, arXiv preprint arXiv:1702.01923 (2017).

\end{thebibliography}

\end{document}